\documentclass[11pt,twoside]{article}

\usepackage{asp2006}
\usepackage{graphicx,natbib}
\bibpunct{(}{)}{;}{a}{}{,}

\begin{document}
\title{On the relationship between magnetic field and mesogranulation}
\author{Alfred~G.~de~Wijn}
\affil{High Altitude Observatory, National Center for Atmospheric Research, P.O.~Box 3000, Boulder, CO 80307, USA, dwijn@ucar.edu}
\author{Daniel M\"uller}
\affil{European Space Agency, Research and Scientific Support Department, c/o NASA Goddard Space Flight Center, Mail Code 671.1, Greenbelt, MD 20771, USA, dmueller@esa.nascom.nasa.gov}

\begin{abstract}
We investigate the relation between Trees of Fragmenting Granules (TFGs) and the locations of concentrated magnetic flux in internetwork areas.
The former have previously been identified with mesogranulation.
While a relationship has been suggested to exist between these features, no direct evidence has yet been provided.
We present some preliminary results that show that concentrated magnetic flux indeed collects on the borders of TFGs.
\end{abstract}

\section{Introduction}

Magnetic field in the quiet solar photosphere, while ubiquitously present, is concentrated at the edges of convective cells.
We refer the interested reader to
	\cite{SSSMF}
for a comprehensive review of small-scale magnetism in the lower solar atmosphere.
The convective flows expunge field from the interiors of granules and concentrate it as ``magnetic elements'' (MEs) in the intergranular downdrafts.
Supergranular flows advect field to supergranular boundaries where it forms the magnetic network.
Gas motions are also expected to push field to the edges of cells of an intermediate, so-called mesogranular scale.
This scale has indeed been observed in the positions of photospheric magnetic elements in internetwork areas
	\citep[e.g.,][]{2003A&A...412L..65D,
	2005A&A...441.1183D,
	2007A&A...462..303T,
	2008ApJ...672.1237L}.
	\cite{1998ApJ...495..973B}
also observed ``voids'' in active network.
However, no evidence has been presented that the observed cells outlined by MEs correspond to mesogranular cells.
Mesogranules have been associated with so-called ``Trees of Fragmenting Granules''
	(\citealp[TFGs,][]{2003A&A...409..299R,
	2004A&A...419..757R},
called ``active granules'' by
	\citealp{2001SoPh..203..211M}).
TFGs consist of repeatedly splitting granules that originate from a single granule.
They may live for several hours, much longer than the lifetime of an individual granule.
Flow fields derived from granular motions also show convergence at the borders of TFGs.
One would thus expect MEs to lie predominantly on TFG boundaries.

\section{Data and Analysis}

\begin{figure*}[tbp]
\begin{center}
\includegraphics[width=130mm]{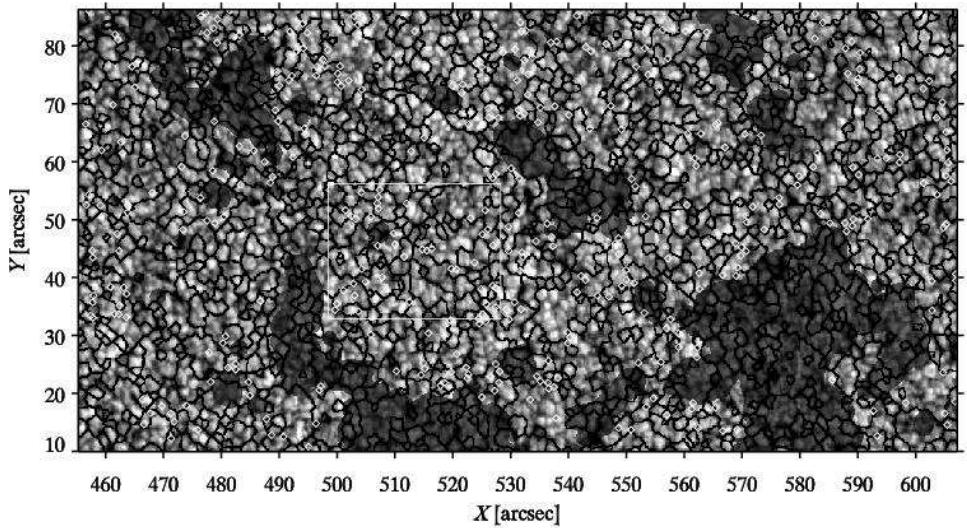}
\caption{Frame number 524 in the Fe~\textsc{i} sequence with the border of TFGs overlaid in black, and the positions of MEs overlaid as white diamonds.
Areas identified as network by
	\cite{2008ApJ...684.1469D}
are shaded dark gray.
The thin white box indicates the area shown in Fig.~\ref{fig:fig2}.
The image has been cropped to remove pixels affected by image motion.}
\label{fig:fig1}
\end{center}
\end{figure*}

An image sequence spanning several hours of moderately quiet sun was recorded using the Solar Optical Telescope on the Hinode spacecraft
	\citep{2007SoPh..243....3K,
	2008SoPh..249..167T,
	2008SoPh..249..197S,
	2008SoPh..249..233I,
	2008SoPh..249..221S}
on March 30, 2007.
The Narrowband Filter Imager was used to record Stokes I and V in the photospheric Fe~\textsc{i} line at 630.2~nm.
More details can be found in
	\cite{2008ApJ...684.1469D}.
Their study of magnetism in quiet-sun internetwork yielded the locations of strong concentrations of magnetic flux that we use in this analysis.

To reduce noise, the Fe~\textsc{i} intensity data is initially filtered using an optimum filter, and averaged over 3 frames in time.
To ensure proper segmentation and separation of granules, the data is scaled up spatially by a factor of two.
Noise is further reduced through convolution with a 8-pixel Gaussian.
We identify granules by computing the curvature $C = 2 I_n - I_{n-1} - I_{n+1}$ in each pixel in 4 directions.
If the minimum curvature is positive, the pixel is labeled as part of a granule.
The granule mask is then extended to values of curvature up to $-2.5\times10^{-4}$, following the CST algorithm
	\citep{2007A&A...471..687R}.
We apply erosion-dilation processing with a 3-pixel +-shaped kernel to remove very small features.
Finally, granules are grouped into families by following them in space and time.
Two granules are considered to be members of the same family if there is a path through the granule mask forward in time that connects them.
At the start of the sequence, all families consist of a single granule.
As time progresses, some families die out, others grow, and new families appear.
If two granules of different families merge, we keep the oldest family.
The lifetime of TFGs is on the order of several hours, so one must expect to wait a similar amount of time before the TFG pattern is established.
Here, we will study a single frame taken about 5 hours after the start of the sequence.

The segmentation of the granules and the subsequent grouping into families has been performed for several variations of the number of frames over which is averaged in time, the width of the Gaussian, and the level to which the masks are extended to negative curvature values.
The chosen parameters appear to be fairly robust, i.e., the resulting pattern does not change much within some range of the chosen values for the parameters.
The process is most sensitive to the amount that the mask is extended.
Too much extension may merge separate granules, causing larger TFGs, while too little extension results in very limited grouping.

\begin{figure*}[tbp]
\begin{center}
\includegraphics[width=100mm]{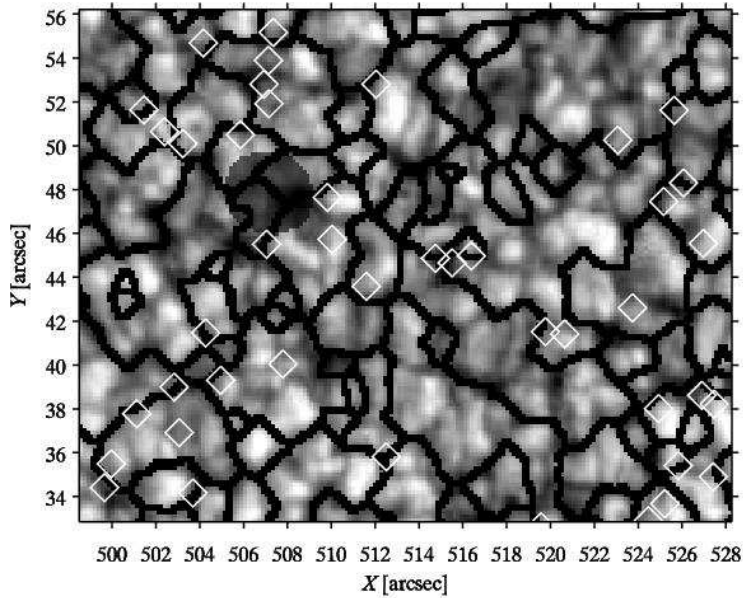}
\caption{Closeup of the white box in Fig.~\ref{fig:fig1}, using the same representation.
The locations of strong flux as indicated by the white diamonds lie close to the borders of TFGs, indicated by thick black lines.}
\label{fig:fig2}
\end{center}
\end{figure*}

\section{Results and Discussion}

Frame number 524 in the Fe~\textsc{i} sequence, recorded at 05:25:33, is shown in Fig.~\ref{fig:fig1}.
It shows the granular pattern, with the borders of TFGs overlaid in black, and the positions of MEs overlaid as white diamonds.
The granular pattern is shaded in dark gray in network areas.
The segmentation there is not good, because of the confusion between granules and bright points.
As a result, the TFG pattern there is not trustworthy.
Also, the analysis by
	\cite{2008ApJ...684.1469D}
did not identify MEs in the network.
This is not a problem for our analysis, because our interest is primarily in the internetwork.

A closeup of the white box is shown in Fig.~\ref{fig:fig2}.
Many MEs appear to lie preferentially near the borders of TFGs.
The interiors of large TFGs are mostly devoid of MEs.
These preliminary results are encouraging.
It thus seems highly likely that the cells described by, e.g.,
	\cite{2005A&A...441.1183D}
and the ``voids'' in maps made with the Hinode SP instrument found by
	\cite{2008ApJ...672.1237L}
correspond to mesogranules.

Further statistical study is required to quantify the relationship between the borders of TFGs and the positions of MEs.
While these preliminary results are encouraging, it must be verified that MEs lie statistically closer to the borders of TFGs than, e.g., to the border of randomly placed cells of similar size.

MEs that emerge in TFG interiors are expected to migrate to the borders in a matter of perhaps an hour, then along the borders of TFGs to the network.
In this context it is of interest to study the motions of MEs in detail and to quantify the contribution of TFGs to the formation of the magnetic network and the diffusion of magnetic field.

\acknowledgements
\emph{Hinode} is a Japanese mission developed and launched by ISAS/JAXA, with NAOJ as domestic partner and NASA and STFC (UK) as international partners.
It is operated by these agencies in co-operation with ESA and NSC (Norway).

\end{document}